# Reynolds-Averaged Turbulence Modeling Using Type I and Type II Machine Learning Frameworks with Deep Learning

Chih-Wei Chang and Nam T. Dinh

Department of Nuclear Engineering
North Carolina State University, Raleigh NC 27695-7909
cchang11@ncsu.edu, ntdinh@ncsu.edu

## Abstract

Deep learning (DL)-based Reynolds stress with its capability to leverage values of large data can be used to close Reynolds-averaged Navier-Stoke (RANS) equations. Type I and Type II machine learning (ML) frameworks are studied to investigate data and flow feature requirements while training DL-based Reynolds stress. The paper presents a method, flow features coverage mapping (FFCM), to quantify the physics coverage of DL-based closures that can be used to examine the sufficiency of training data points as well as input flow features for data-driven turbulence models. Three case studies are formulated to demonstrate the properties of Type I and Type II ML. The first case indicates that errors of RANS equations with DL-based Reynolds stress by Type I ML are accumulated along with the simulation time when training data do not sufficiently cover transient details. The second case uses Type I ML to show that DL can figure out time history of flow transients from data sampled at various times. The case study also shows that the necessary and sufficient flow features of DL-based closures are first-order spatial derivatives of velocity fields. The last case demonstrates the limitation of Type II ML for unsteady flow simulation. Type II ML requires initial conditions to be sufficiently close to reference data. Then reference data can be used to improve RANS simulation.

## 1. Introduction

Reynolds-averaged Navier-Stokes (RANS) equations are widely used in fluid engineering simulation and analysis due to its computational efficiency. Reynolds stress is essential to close RANS equations. Linear eddy viscosity models (LEVMs) are attractive to represent Reynolds stress due to their computational efficiency. LEVMs include Spalart-Allmaras [1], k-ε [2], and k-ω [3] models that require extensively evaluated and calibrated for different flow characteristics. Consequently, performance of different models is limited in their calibration domains and exhibit different degrees of uncertainty in prediction. Tracey, Duraisamy & Alonso [4] demonstrated that the Menter's k-ω model [5] yielded large uncertainty for the calculation of Reynolds stress anisotropy.

The growing interest in machine learning (ML), especially deep learning (DL), application for science and engineering leads to data-driven modeling of Reynolds stress. Deep learning (DL) [6] belongs to a branch of machine learning (ML), and it is a universal approximator [7] that can capture underlying correlations behind data. DL (or deep neutral networks, DNNs) with its hierarchical model structure can leverage values of large datasets from relevant experiments and simulations without a limitation to a single data source. Such a feature can achieve total data-model integration [8] that is capable of constructing fluid closures over a range of flow regimes. Based on data and knowledge requirements, Chang & Dinh [9] classified five types of ML frameworks for using ML in thermal fluid simulation. The present work employs Type I (physics-separated) and Type II (physics-evaluated) ML frameworks [9] for the development of DL-based Reynolds stress.

Type I ML [9] requires a scale separation assumption [10, 11] such that closure relations can be derived separately from conservation equations using experimental data and ML models. Then simulation solves conservation equations with embedded ML-based closures. Closure relations by Type I ML are iteratively queried during simulation. Previous Type I ML applications included system-level flow modeling and Reynolds-averaged turbulence modeling. Chang & Dinh [12, 13] employed DL-based closures to model system pressure drop and boiling channel flow. Ma, Lu & Tryggvason [14] used neural networks (NNs) to surrogate fluid closures from simulating isothermal bubbly flow. Tracy, Duraisamy & Alonso [15] and Zhang & Duraisamy [16] used shallow NNs to achieve data-driven turbulence modeling. Although Type I ML has been employed for flow simulation, previous works do not investigate in data requirement for developing DL-based closures with predictive capabilities. In the present work, we demonstrate a method to quantify the predictive capability of DNNs based on training datasets with different qualities.

Type II ML [9] requires knowledge on selections of simulation models as low-fidelity models, which are efficient for computation. Model uncertainty can be reduced by high-fidelity simulation such as direct numerical simulation and large eddy simulation. Closure relations by Type II ML can be built to correlate the inputs (mean flow properties by low-fidelity simulation) and targets (quantities of interest by high-fidelity simulation). To employ Type II ML in thermal fluid simulation, we need to run low-fidelity simulation with embedded mechanistic closures to obtain mean flow properties as model inputs. Then we use the inputs to query outputs from ML-based closures. The outputs, fixed values, are implemented in the low-fidelity model to replace the mechanistic closure. The Type II ML approach is similar to the strategy to leverage high-fidelity data proposed by Lewis *et al.* [17]. Previous Type II ML works included Reynolds-averaged turbulence modeling. Ling, Kurzawski & Templeton [18] used DL-based Reynolds stress with embedded Galilean invariance to close RANS, and they demonstrated predictive capabilities of the DL-based stress for flows in different geometries. However, they focused on steady flow applications. In the percent work, we demonstrate that DL-based Reynolds stress can be used for unsteady flow simulation.

Kutz [19] addressed several open challenges for applications of DL-based closures such as requirements of training data. Ling & Templeton [20] applied the Mahalanobis to indicate the similarity between training and testing data. In the present paper, Type I and Type II ML [9] are studied to investigate the requirements of DL-based Reynolds stress development. The case study in the present work is formulated based on: (i) How large training datasets should be to train DL-based closures? (ii) What are the necessary and sufficient flow features? These questions are fundamental to define data requirements of thermal fluid simulation with embedded DL-based closures. We define the RANS simulation with embedded DL-based closures as RANS-DL. The present paper investigates how to apply RANS-DL to accomplish data-driven turbulence modeling of computational fluid dynamics by assimilating available, relevant, and adequately evaluated data.

The paper is structured to include: Assumption testing (section 2), Formulation of the case study (section 3), Flow features coverage mapping (section 4), Implementation of ML frameworks (section 5), Results (section 6), Lessons learned (section 7), and Conclusions (section 8).

## 2. Assumption testing

Data-driven modeling by DL requires a substantial amount of data. To investigate the data requirement, we formulate three tests to investigate how to use DL to close RANS equations. The first test is to find the essential datasets to reconstruct the history of flow transients by RANS-DL. The second test is to determine necessary flow features as inputs for DL-based closures. Finally, the last test is to examine the applicability of Type I and Type II ML for unsteady flow simulation.

### 2.1. Assumption testing on the data requirement

We assume that DL can discover hidden time derivatives from spatially distributed velocity fields collected from different flow patterns. Therefore, we can sample data from various simulation time steps and train DNNs using total data. The assumption testing includes training data obtained from reference solutions by RANS simulation. The success criterion depends on whether RANS-DL can reconstruct reference solutions.

### 2.2. Assumption testing on the flow feature selection

We assume that the sufficient and necessary flow features can be defined by spatial derivatives of velocity fields. DL belongs to supervised learning [21] which requires inputs and targets for training. For DL-based Reynolds stress, the inputs are flow features that represent mean flow properties, and the target is the Reynolds stress tensor. We select input flow features based on the incompressible momentum equation [22] given by Eq. (1) where $\bar{u}$ is the mean velocity and $i$, $j$, and $k$ denote directions. $D/Dt$, $\rho$, $\bar{p}$, $\bar{\tau}_{ij}$, $\mu$, and $\delta_{ij}$ are the material derivative, fluid density, mean pressure, Reynolds stress tensor, molecular viscosity, and Kronecker delta. We can manipulate Eq. (1) into Eq. (2) that shows the dependency of Reynolds stress. Eq. (3) gives the derivative of Reynolds stress as a function of several bases from Eq. (2).

$$\rho \frac{D\bar{u}_i}{Dt} = -\frac{\partial \bar{p}}{\partial x_j} + \frac{\partial}{\partial x_j}\left[\mu\left(\frac{\partial \bar{u}_i}{\partial x_j} + \frac{\partial \bar{u}_j}{\partial x_i}\right) - \frac{2}{3}\mu\delta_{ij}\frac{\partial \bar{u}_k}{\partial x_k}\right] + \frac{\partial \bar{\tau}_{ij}}{\partial x_j} \tag{1}$$

$$\frac{\partial \bar{\tau}_{ij}}{\partial x_j} = \rho \frac{D\bar{u}_i}{Dt} + \frac{\partial \bar{p}}{\partial x_j} - \frac{\partial}{\partial x_j}\left[\mu\left(\frac{\partial \bar{u}_i}{\partial x_j} + \frac{\partial \bar{u}_j}{\partial x_i}\right) - \frac{2}{3}\mu\delta_{ij}\frac{\partial \bar{u}_k}{\partial x_k}\right] \tag{2}$$

$$\frac{\partial \bar{\tau}_{ij}}{\partial x_j} = f\left(\rho \frac{D\bar{u}_i}{Dt}, \frac{\partial \bar{p}}{\partial x_j}, \frac{\partial}{\partial x_j}\left(\mu\frac{\partial \bar{u}_i}{\partial x_j}\right), \frac{\partial}{\partial x_j}\left(\mu\frac{\partial \bar{u}_j}{\partial x_i}\right), \frac{\partial}{\partial x_j}\left(\mu\delta_{ij}\frac{\partial \bar{u}_k}{\partial x_k}\right)\right) \tag{3}$$

Based on the first assumption testing, time derivatives are not selected as training inputs since data are steady for each dataset. The merged PISO-SIMPLE algorithm [23, 24] is implemented for pimpleFoam that use the projection method [25, 26] to solve RANS equations. Since pressure is separately solved from the momentum equation, we further assume that the pressure term can be excluded from training inputs. As a result, the essential flow features for DL-based Reynolds stress can be represented by remaining spatial derivatives of velocities. Eq. (4) uses the matrix form to show Reynolds stress ($\tau$) as a dyadic product between the derivative operator and velocity ($\mathbf{U}$). The dyadic product in Eq. (4) results in nine velocity derivatives as input flow features. Targets are the Reynolds stress symmetry tensor that includes six stress components. The assumption testing is to examine whether those nine flow features are sufficient and necessary for surrogating Reynolds stress by DNNs.

$$\boldsymbol{\tau} = f\left(\left(\nabla \otimes \mathbf{U}\right)^T\right) \tag{4}$$

### 2.3. Assumption testing on Type I and Type II ML

Type I and Type II ML can both build closure relations for thermal fluid simulation. We implement these two frameworks for data-driven turbulence modeling using DL-based Reynolds stress in section 5 to explore performance of each framework. Closure relations are iteratively queried in Type I ML while solving conservation equations. Type II ML solves conservation equations with fixed closure relations. Therefore, we assume Type I ML is capable of simulating unsteady flow while Type II ML has limitations in transient problems. The goal of Type I ML is to reproduce transient solutions by RANS. Type II ML aims at exploring what the magnitude of errors can be before it is too late to bring solutions to the quasi-steady state from a transient state. The assumption testing is to evaluate whether the goals for each ML framework is achieved.

## 3. Formulation of the case study

### 3.1. Numerical experiment

The numerical experiment is formulated to evaluate performance of RANS simulation with embedded DL-based Reynold stress (RANS-DL). The RANS simulation using the k-ε model serves as reference solutions that are used to train DL-based Reynolds stress. Fig. 1 depicts the simulation configuration created by Pitz and Daily [27] which is used to explore the requirements for data-driven turbulence modeling. The 2D geometry includes a backward-facing step and converging nozzle. System characteristics are summarized in Table 1. This geometry configuration is complex enough since unsteady flow is affected by the turbulence mixing layer growth, entrainment rate, and reattachment length. The k-ε model has been validated [28] for this geometry. The pimpleFoam solver [29] in OpenFOAM [30] is used to generate data for the development of DL-based Reynolds stress.

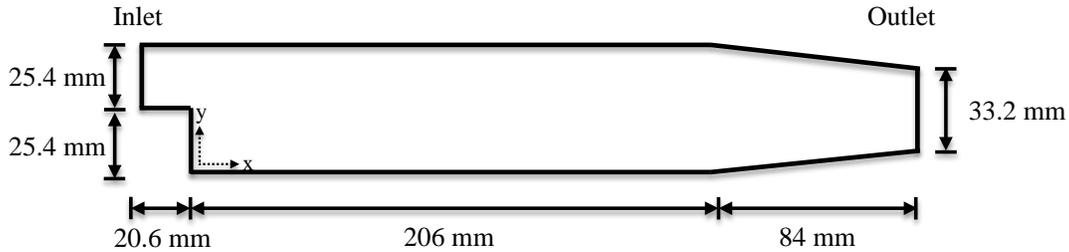

Fig. 1. Geometry configurations of RANS simulation.

Table 1. System characteristics for RANS simulation.

| Initial Conditions | |
|---|---|
| Velocity | 0 m/s |
| Pressure | 0 bar |
| Boundary Conditions | |
| Inlet velocity | (10, 0, 0) m/s |
| Inlet pressure | Zero gradient |
| Outlet velocity | Zero gradient |
| Outlet pressure | 0 bar |
| Transport Properties | |
| Kinematic viscosity | $10^{-5}$ m$^2$/s |

### 3.2. Training data

Training data are generated by RANS simulation with the k-ε model. The equations are solved by pimpleFoam using fixed time step, 2.4x10$^{-5}$ sec. Four datasets are created and listed in Table 2.

The first three datasets involve millions of data points, and the last dataset has hundreds of thousands of data points. The data in T10A and T10B are uniformly sampled from ten various times, and sampling time ranges are given in Table 2. T10A includes less transient details than T10B because the data are sampled from a coarse time interval in T10A. The baseline dataset includes data sampled from twenty separate times, and it is used to evaluate performances of RANS-DL.

Table 2. Generated datasets sampled at various times with distinct flow patterns.

| Dataset | Data Quantity | | Total datasets sampled from various times (sec) | Description |
|---|---|---|---|---|
| | Inputs (x$10^6$) | Targets (x$10^5$) | | |
| T10A | 1.15 | 7.65 | [0.01, 0.1] | Training dataset |
| T10B | 1.15 | 7.65 | [0.010024, 0.01024] | Training dataset |
| Baseline | 2.30 | 15.30 | [0.010024, 0.01048] | Validating dataset |
| QSS | 0.11 | 0.76 | 1 | Validating dataset |

The QSS (quasi-steady-state) dataset is sampled from RANS simulation. The QSS solution is checked by the mean square error (MSE) defined by Eq. (5) where $N$, $i$, $y$, and $y_{ref}$ are the total data points, $i^{th}$ data point, solution at the current time step ($t^n$) and previous time step ($t^{n-1}$). Fig. 2 depicts MSE analysis for the simulation running from 0.1 to 1 sec. Based on the result, the QSS dataset is sampled at $t = 1$ sec.

$$MSE = \frac{\sum_{i=1}^{N}(y_i - y_{ref,i})^2}{N} \tag{5}$$

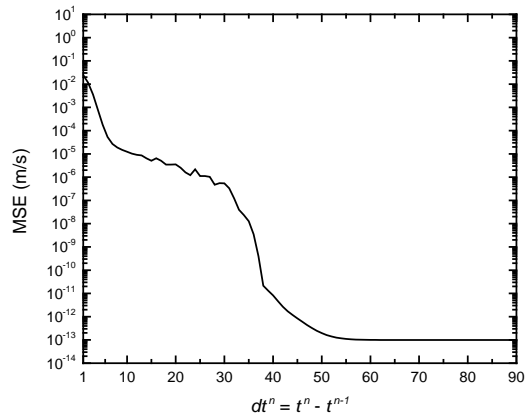

Fig. 2. MSE analysis to check whether the quasi-steady-state (QSS) condition is achieved.

## 4. Flow features coverage mapping (FFCM)

Flow features coverage mapping (FFCM) is a 2D graph that shows distributions of high-dimensional flow features. It can be used to quantify whether the physics is sufficiently covered by training data. If the mapping between training and applications shows similar distributions, we are confident in predictive results by RANS-DL. The discrepancy between two FFCM can be quantified by Eq. (5) using MSE for point-by-point comparisons.

FFCM is obtained by a two-step approach. First, k-mean clustering [31, 32] is employed to classify flow features based on their similarities by computing distances between centroids of clusters and data points. Data points are assigned to a cluster if the minimum distance is achieved. Then centroids of clusters are updated based on data points within a cluster. The process is iteratively repeated until convergence is reached. The clustering results are multidimensional because our selected flow features include nine components of spatial derivatives of velocity fields.

Second, we use t-SNE (t-distributed stochastic neighbor embedding) [33] to visualize the clustering result that is flow features coverage mapping. t-SNE is a method for dimensionality reduction, and it can project high-dimensional data in a 2D or 3D graph while preserving characteristics of data points. t-SNE first calculates pairwise conditional probabilities using Gaussian kernels for high-dimensional data such that similar points have high probabilities while dissimilar points have low probabilities. Then t-SNE uses a $t$ distribution to measure pairwise similarities of low-dimensional data points. Positions of low-dimensional points are calculated by minimizing Kullback-Leibler divergence [34] between $t$ and Gaussian distributions in low-dimensional and high-dimensional spaces. A $t$ distribution has fat tails at both ends that ensure dissimilar points in low-dimensional space to be placed away from similar points. Therefore, t-SNE can embed high-dimensional data in a low-dimensional space. By k-mean clustering and t-SNE, we can build FFCM to quantify similarities of flow features between RANS-DL and training datasets. FFCM can be used to evaluate whether the training of DL-based Reynolds stress is sufficient.

Fig. 3 depicts FFCM by T10A data at two times. The discrepancy can be quantified by computing the Euclidean distance. Eq. (6) gives the mean Euclidean distance ($d$) where $N$ and $i$ denote the total data and $i^{th}$ data point. $(x, y)$ and $(\hat{x}, \hat{y})$ are coordinates from two distinct mapping. We can observe significant differences between Fig. 3(a) and Fig. 3(b), and their distance is 43.67. Fig. 4 shows FFCM by T10B data at two times, and the distance between Fig. 4(a) and Fig. 4(b) is 11.31. The results indicate that a sharp transient happens between two consecutive time intervals in T10A. It is because the sampling interval is coarser in T10A than the interval in T10B. Fig. 3 and Fig. 4 serve as the references that are used to examine physics coverages of DL-based Reynolds stress in applications. If features mapping in applications has a similar distribution as the references, we expect a good prediction of velocity fields by RANS-DL.

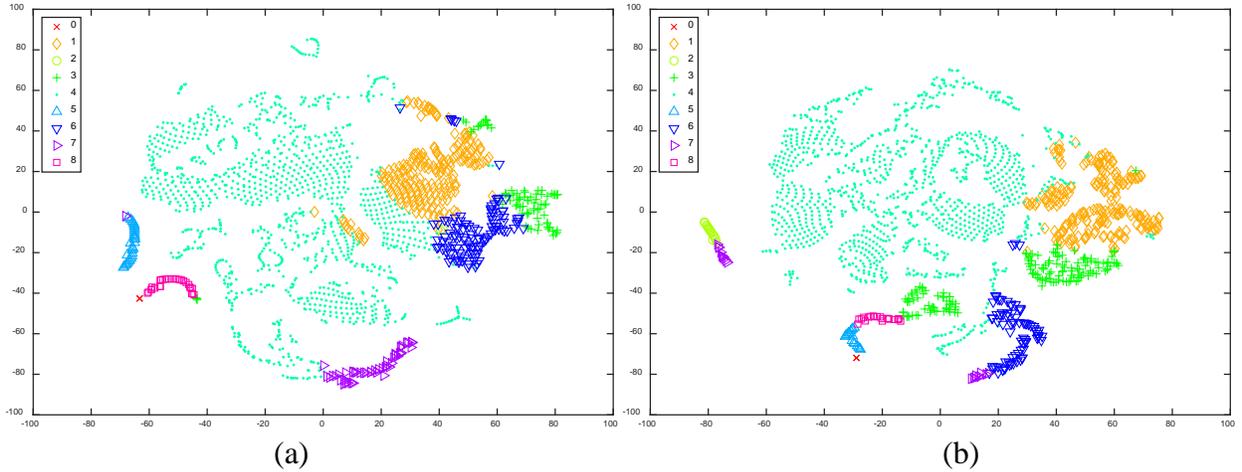

Fig. 3. Visualization of flow features coverage mapping (FFCM) using t-SNE at (a) t = 0.01 sec and (b) t = 0.02 sec from T10A dataset. The flow features are clustered by k-means clustering with variously labeled colors.

$$d = \sqrt{\frac{1}{N}\sum_{i=1}^{N}\left[(x_i - \hat{x}_i)^2 + (y_i - \hat{y}_i)^2\right]} \qquad (6)$$

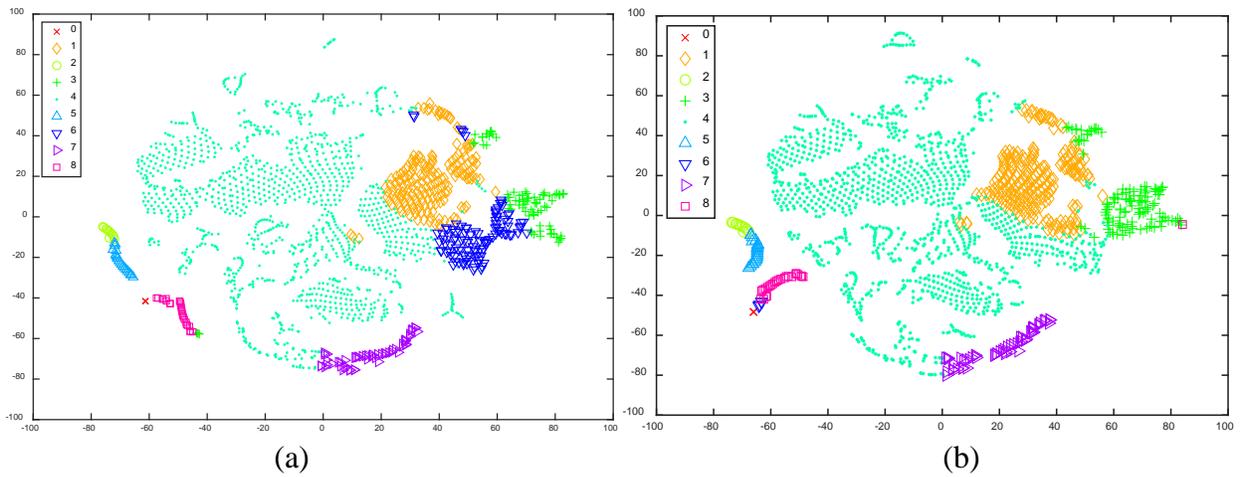

Fig. 4. Visualization of flow features coverage mapping (FFCM) using t-SNE at (a) t = 0.010096 sec from T10B dataset and (b) t = 0.010456 sec from the V10 dataset. The flow features are clustered by k-means clustering with variously labeled colors.

## 5. Implementation of ML frameworks

### 5.1. Implementation of DL-based Reynolds stress

We use DL (or DNNs) to surrogate Reynolds stress due to the nonparametric modeling feature of DNNs. This feature allows the model form of DNNs to be adaptive based on various data quantities. Fig. 5 depicts a structure of DNNs including nine input flow features and six output Reynolds stress components. We use Tensorflow [35] to design a ten-layer DNN with 512 hidden units (HUs) for DL-based Reynolds stress. The activation function ($\sigma$), rectified linear units (ReLU) [36], is used in each hidden layer (HL). Eq. (7) defines a cost function by Euclidean loss where $N$, $y_{i,data}$, and $y_{i,model}$ are the total number of training data, $i^{th}$ training data, and $i^{th}$ model solution. DNNs' parameters such as weights and biases are tuned based on data using Adam [37] algorithm.

$$E = \frac{1}{2N}\left[\sum_{i=1}^{N}\left(y_{i,\text{model}} - y_{i,\text{data}}\right)^2\right] \tag{7}$$

Large data can increase the difficulty of training DNNs. In fact, neural networks with many HLs may suffer from gradient vanishing or explosion issues that slow the learning process. Batch normalization (BN) [38] is a method that can reduce internal covariate shifts in DNNs to prevent those issues. Therefore, BN is implemented in each HL to accelerate the speed of training. Fig. 6(a) depicts the comparison of Euclidean loss by training DNNs with different data. T10A010 denotes that training data are from T10A at $t = 0.1$ sec while T10A includes data from all times. T10A010 only involves one-tenth data points of T10A. Fig. 6(a) shows that the learning using T10A is much slower than T10A010. Fig. 6(b) reveals that the learning becomes fast while implementing BN in DNNs. Fig. 7 depicts a model-data plot to show that DL-based Reynolds stress is well-trained by T10B dataset since model outputs agree with data.

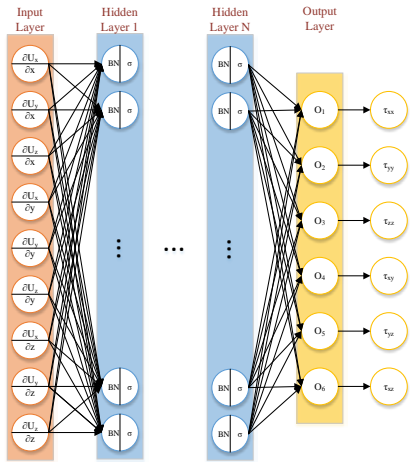

Fig. 5. Structure of DL-based Reynolds stress with batch normalization (BN).

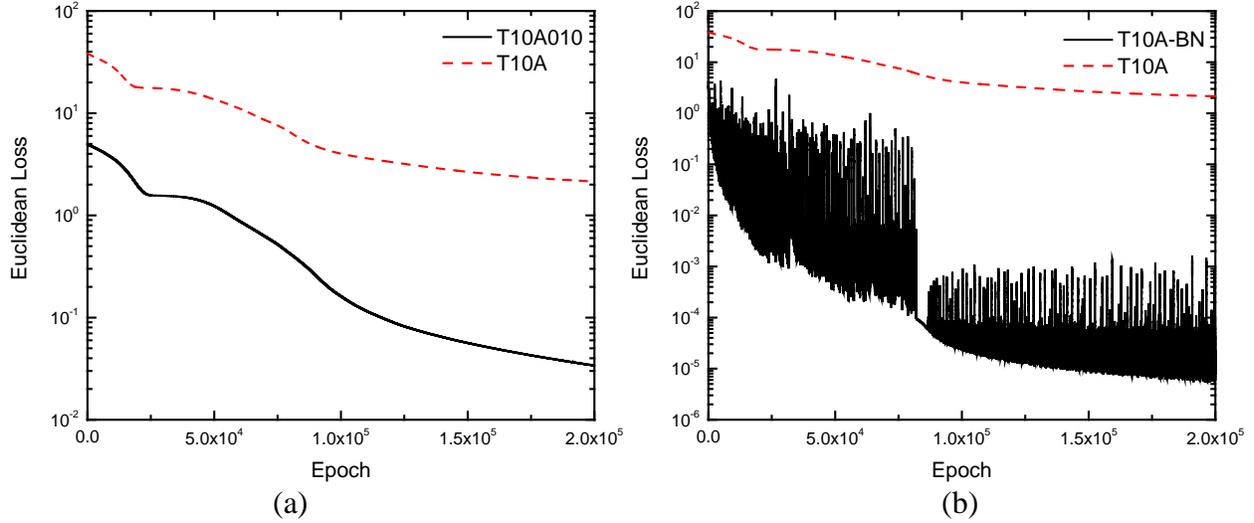

Fig. 6. (a) Comparison of Euclidean loss between DNNs using training datasets, T10A010 and T10A. (b) Comparison of Euclidean loss between DNNs using training datasets, T10A and T10A-BN.

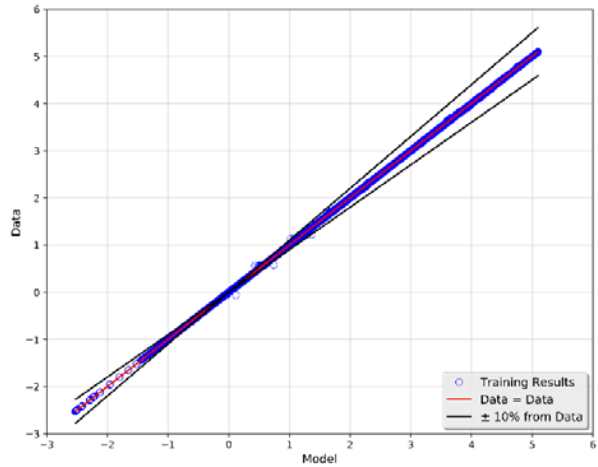

Fig. 7. Model-data plot for visualizing the performance of DL-based Reynolds stress trained by T10B dataset.

## 5.2. Implementation of Type I ML for data-driven turbulence modeling

The goal of Type I ML [9] is to build DL-based Reynolds stress that allows RANS-DL to reconstruct the results by baseline solutions. Fig. 8 depicts Type I ML framework for the development of DL-based Reynolds stress. The procedure includes the following elements:

*Element 1.* Assume the separation of scales is achievable such that Reynolds stress can be calculated from RANS data ($\Psi_{RANS}$) using Boussinesq hypothesis with the $k$-$\omega$ model. Transient data ($\Psi_{RANS}$) are given in Table 2.

*Element 2*. Compute a dyadic product between the gradient operator ($\nabla$) and velocity fields (**U**) from $\Psi_{RANS}$ that results in nine velocity derivatives as flow features (**Q**).

*Element 3*. Select flow features (**Q**) calculated by element 2 as training inputs for element 5.

*Element 4*. Compute training targets, Reynolds stress ($\tau$), by solving the linear eddy viscosity model using the velocity fields, turbulence kinetic energy, and dissipation rate from $\Psi_{RANS}$. The results become targets for element 5.

*Element 5*. Utilize Adam algorithm [37] to capture underlying correlations between flow features (**Q**) and Reynolds stress ($\tau$) by DNNs. After the training, output the DL-based Reynolds stress, $DNN(Q(\Psi_{RANS}))$, to the next element.

*Element 6*. Constrain the output of $DNN(Q(\Psi_{RANS}))$ by $g(DNN(Q(\Psi_{RANS})))$ to satisfy the property of 2D simulations, i.e., Reynolds stress components should be zero in x-z and y-z directions.

*Element 7*. Implement DL-based Reynolds stress in pimpleFoam solver. Then solve the RANS with the embedded DL-based closure that is iteratively queried. The baseline dataset in Table 2 is used to evaluate the performance of RANS-DL.

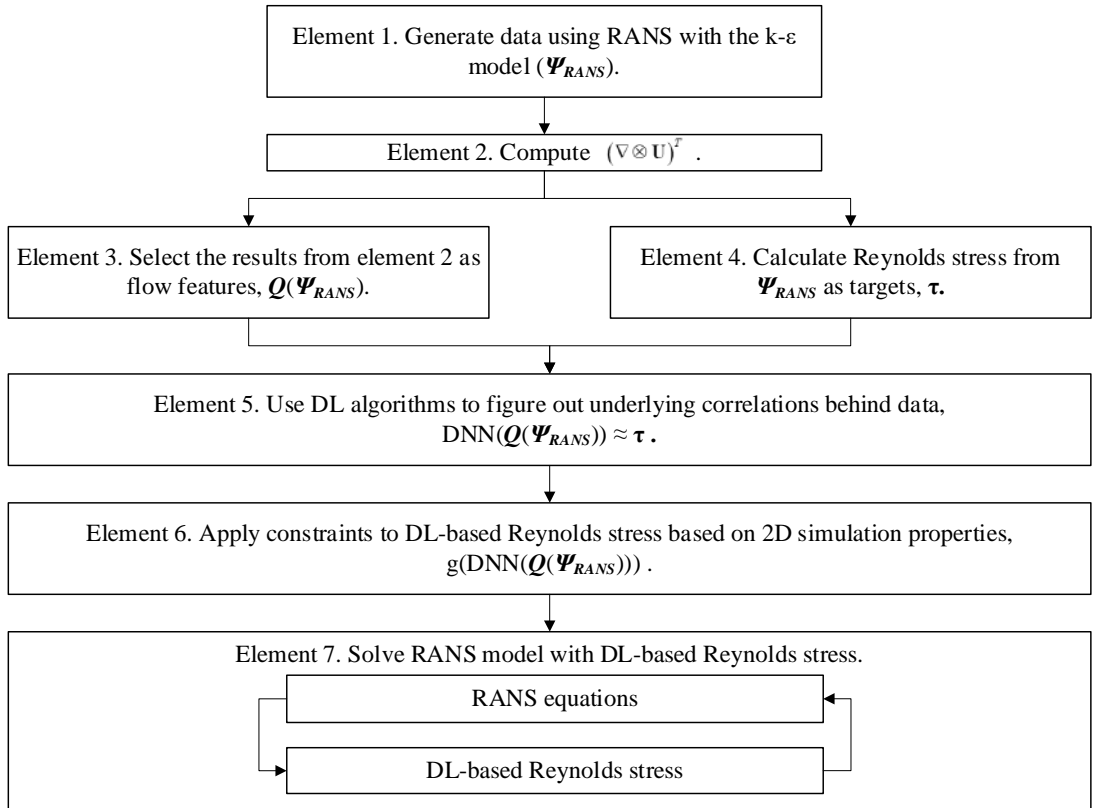

Fig. 8. Type I ML for data-driven turbulence modeling using RANS equations with DL-based Reynolds stress.

### 5.3. Implementation of Type II ML for data-driven turbulence modeling

The goal of Type II ML [9] is to use the reference Reynolds stress to bring solutions to the quasi-steady state (QSS) from various transient states. The reference Reynolds stress is calculated by QSS dataset in Table 2. Fig. 9 depicts the workflow of Type II ML for data-driven turbulence modeling. The procedure involves the following elements:

*Element 1*. Solve RANS equations with the $k$-$\varepsilon$ model until the solution ($\Psi_{RANS,\,\omega}$) reaches the quasi-steady state. QSS dataset is given in Table 2, and it serves as the reference that can be used to compute raining targets and to evaluate whether RANS-DL achieves the goal. The goal is to test if Type II ML can bring solutions from various transient states to the quasi-steady state.

*Element 2*. Perform RANS simulations with the $k$-$\varepsilon$ model to obtain solutions ($\Psi_{RANS}$) at various transient states.

*Element 3*. Compute a dyadic product between the gradient operator ($\nabla$) and velocity fields (**U**) from $\Psi_{RANS}$. The results include nine velocity derivatives.

*Element 4*. Select the nine spatial velocity derivatives from element 3 as flow features (**Q**) which become training inputs for element 6.

*Element 5*. Compute reference Reynolds stress ($\tau$) by Boussinesq hypothesis with the $k$-$\varepsilon$ model and $\Psi_{RANS,\,\omega}$. The results become targets for element 6 that can supervise DL algorithms to learn from data.

*Element 6*. Utilize DL to correlate flow features (**Q**) from various transient states to the reference Reynolds stress ($\tau$). After the training, output DL-based Reynolds stress, $DNN(\mathbf{Q}(\Psi_{RANS}))$, to element 8.

*Element 7*. Execute RANS simulations with the $k$-$\varepsilon$ model ($\Psi'_{RANS}$), and stop simulations at a particular transient state. Then use the solution to compute new flow features (**Q'**) as inputs to element 8.

*Element 8*. Query the values of DL-based Reynolds stress by new flow features (**Q'**). Then output fixed Reynolds stress fields to element 9.

*Element 9*. Implement fixed Reynold stress fields in pimpleFoam solver to close RANS equations.

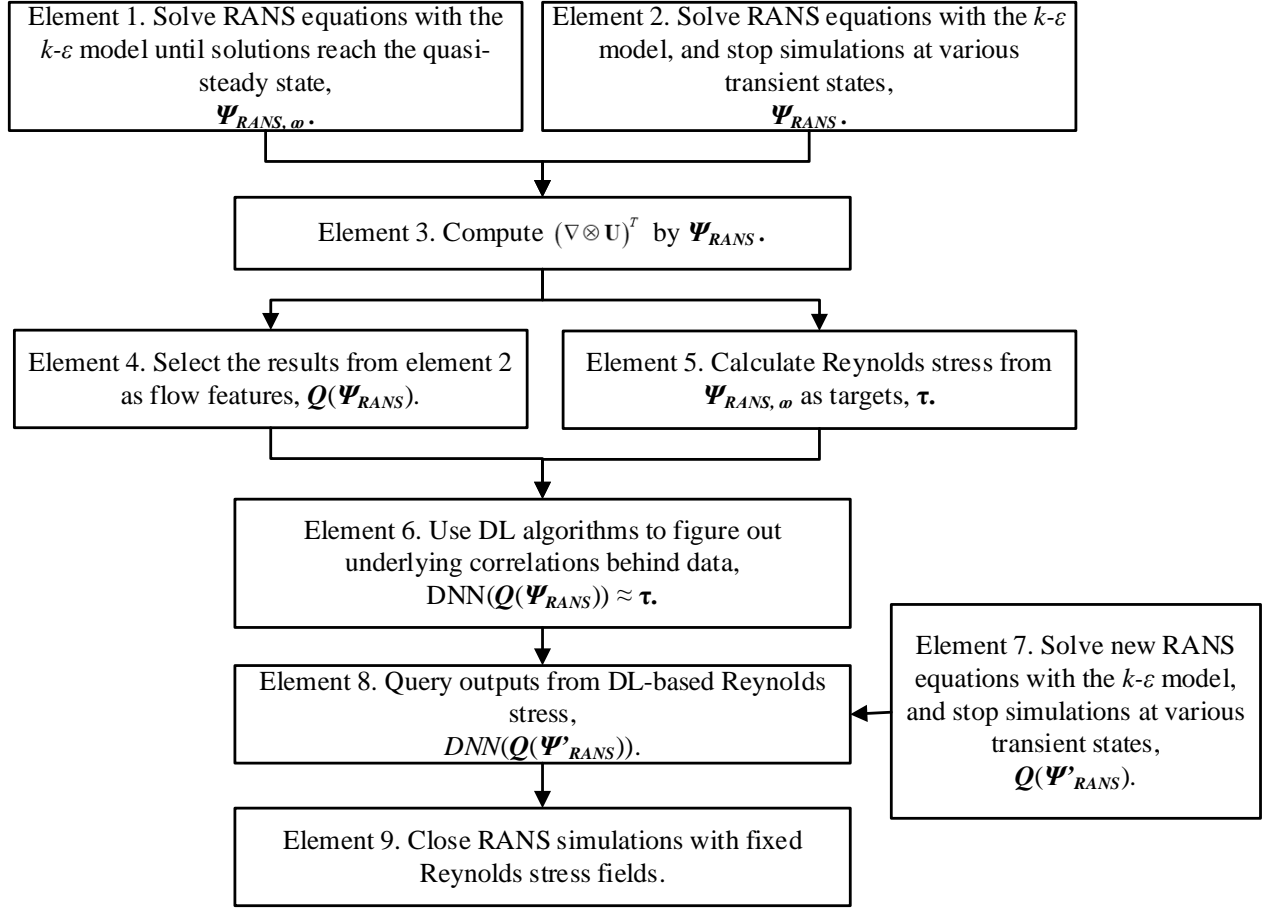

Fig. 9. Type II ML for data-driven turbulence modeling using RANS equations with DL-based Reynolds stress.

Table 2 shows that Type II ML only includes one-tenth data points of the data used in Type I ML. Since Fig. 7 demonstrates that DL can successfully infer a surrogate that fits large data by T10B, the challenge of Type II ML is not subject to performance of DL. Instead, the challenge is whether Type-II ML can bring solutions to the quasi-steady state from an arbitrary transient state. To investigate this limitation, we directly explore the problem from element 7. We assume that DNNs can output ideal fields of reference Reynolds stress without uncertainty. No matter what flow features are inputted, DL-based Reynolds stress can always deliver the reference stress field. Therefore, we can implement the reference stress field in RANS equations and evaluate the performance of Type II ML while simulating unsteady flow.

## 6. Results

To explore the assumption testing, we analyze results into three sections. The first section shows that errors of RANS-DL by Type I ML are accumulated along with the simulation time. The second section focuses on testing whether RANS-DL by Type I ML can recover the baseline solutions for unsteady flow. The last section aims at testing if RANS-DL by Type II ML can find solutions to the quasi-steady state from a transient state.

### 6.1. Error accumulation along with the time during simulation

The case is formulated to analyze how errors propagate when training data do not sufficiently cover the flow features in applications. We use T10A to train DL-based Reynolds stress, and implement the stress in RANS equations. Then the simulation is started at $t = 0.1$ sec using initial conditions obtained from the baseline. Fig. 10(a) depicts velocity profiles at $t = 0.015$ sec by RANS-T10A001 which stands for RANS-DL starting at $t = 0.01$ sec. The trend of RANS-T10A001 velocity (dash line) agrees with the baseline (solid line). At $t = 0.065$ sec, Fig. 10(b) shows that the uncertainty of RANS-T10A006 (dash-dot line) is much smaller than the uncertainty of RANS-T10A001 while comparing results to the baseline. RANS-T10A006 represents RANS-DL starting at $t = 0.06$ sec. Fig. 10 indicates that T10A does not contain enough data to allow DL to capture all transient behaviors. Initially, there are strong transients in unsteady flow simulation, and errors caused by DL-based Reynolds stress grow along with the time.

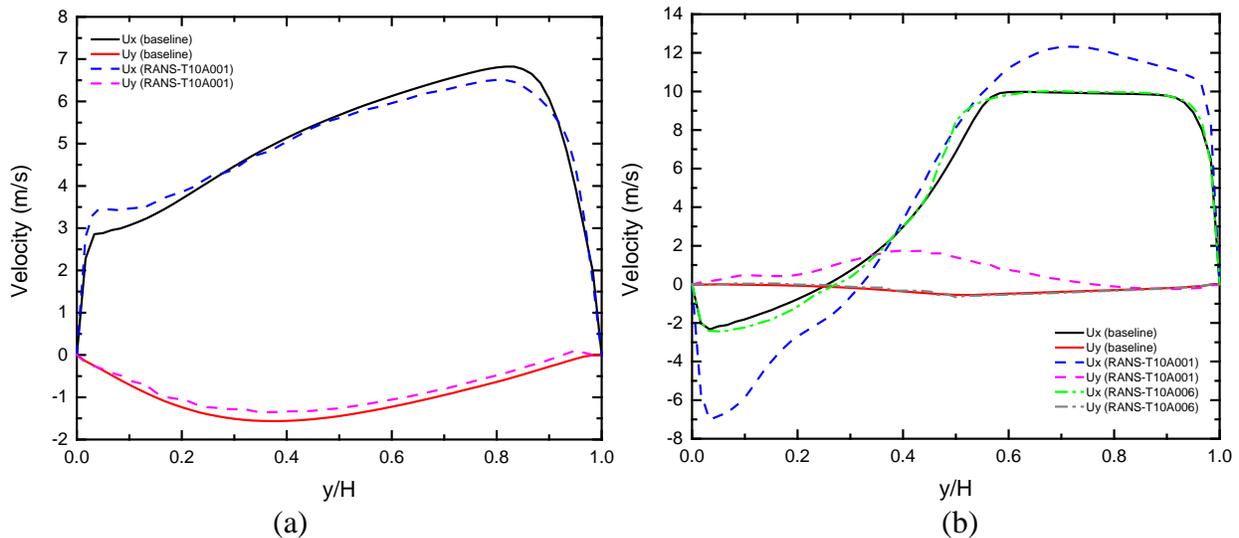

Fig. 10. (a) Comparison of velocities between the baseline and RANS-T10A at x = 0.07 m and t = 0.015 sec with initial conditions from the baseline at t = 0.01 sec. (b) Comparison of velocities of the baseline, RANS-T10A001, and RANS-T10A006 at x = 0.07 m and t = 0.065 sec with initial conditions from the baseline at t = 0.01 and 0.06 sec for RANS-T10A001 and RANS-T10A006.

## 6.2. Exploration of data requirements to reconstruct the RANS solution

This task is formulated to compare the performance of RANS-DL with the stress closures trained by T10A and T10B. Fig. 11 depicts initial Reynolds stress and velocities of the baseline, RANS-T10A, and RANS-T10B at $t = 0.01$ sec.

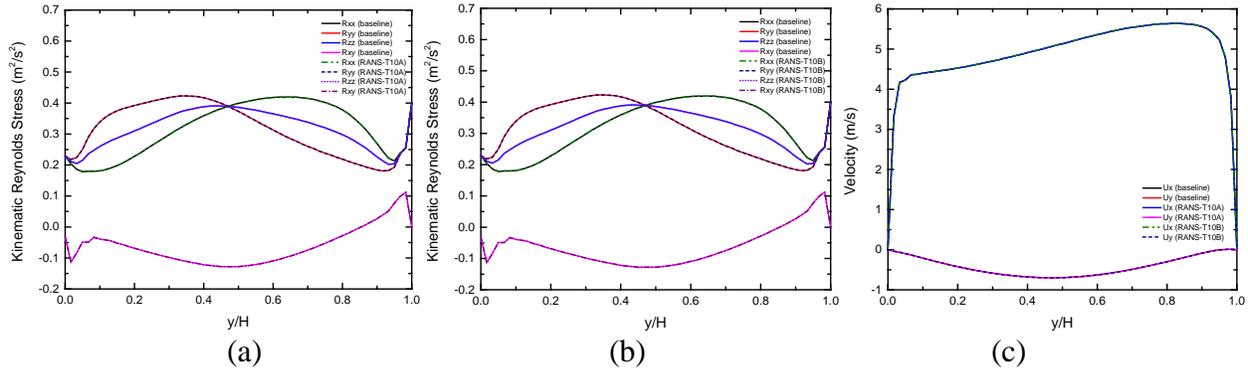

Fig. 11. Comparison of initial kinematic Reynolds stress (a) between the baseline and RANS-T10A (b) and between the baseline and RANS-T10B at t = 0.01 sec. (c) Comparison of the initial velocities for the baseline, RANS-T10A, and RANS-T10B at t = 0.01 sec.

Fig. 12-Fig. 15 illustrate the results by RANS-DL at $t = 0.01012, 0.01024, 0.01036,$ and $0.01048$ sec. The first two times are within the training domain of T10B while the last two times are in extrapolation domains. For T10A, all simulation times are in extrapolation domains because its data are sampled from a coarse time interval. Therefore, RANS simulation using DL-based Reynolds stress by T10A (RANS-T10A) yields large uncertainty than RANS-T10B. Fig. 14(b) shows that RANS-T10B starts to deviate from the baseline when the simulation is outside of the training domain. Although the simulation time is too short to make significant changes in velocity profiles, Fig. 15(c) depicts that the velocity of RANS-T10A is different from the baseline at the bottom location.

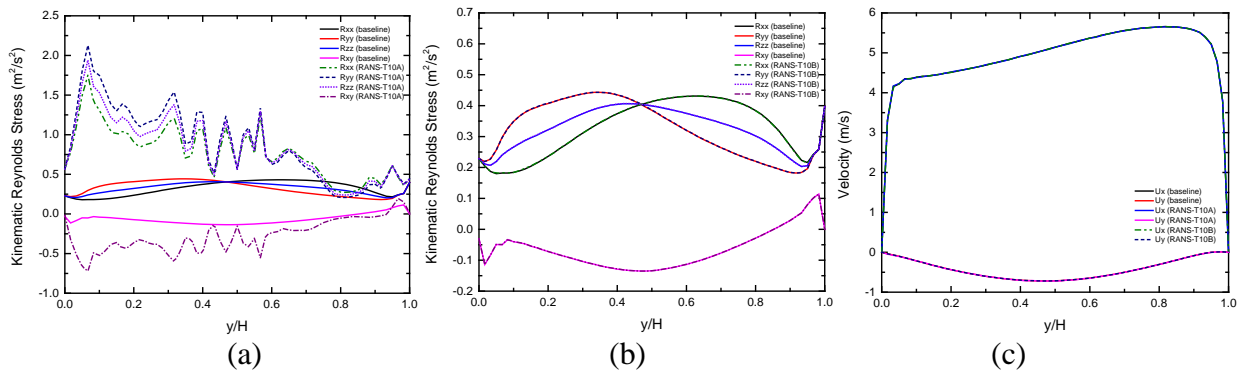

Fig. 12. Comparison of kinematic Reynolds stress (a) between the baseline and RANS-T10A and (b) between the baseline and RANS-T10B at t = 0.01012 sec and x = 0.07 m. (c) Comparison of the velocity of the baseline, RANS-T10A, and RANS-T10B.

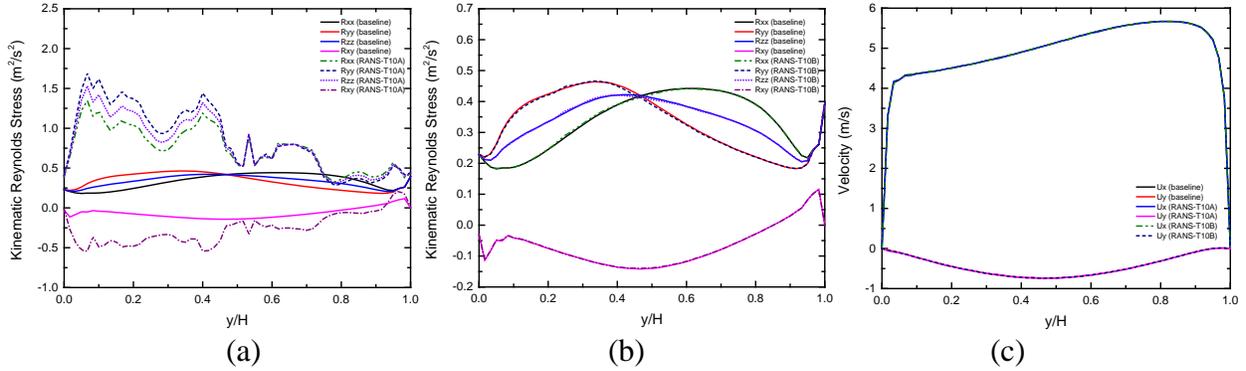

Fig. 13. Comparison of kinematic Reynolds stress (a) between the baseline and RANS-T10A and (b) between the baseline and RANS-T10B at t = 0.01024 sec and x = 0.07 m. (c) Comparison of the velocity of the baseline, RANS-T10A, and RANS-T10B.

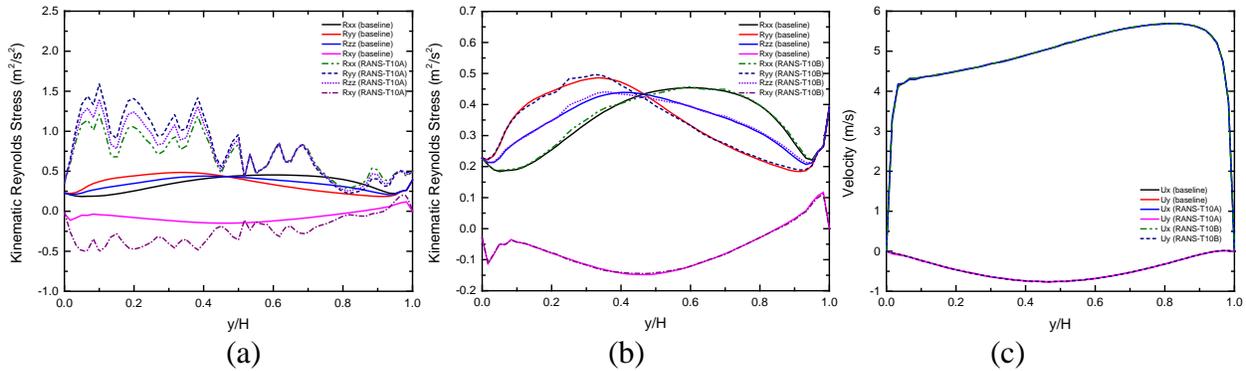

Fig. 14. Comparison of kinematic Reynolds stress (a) between the baseline and RANS-T10A and (b) between the baseline and RANS-T10B at t = 0.01036 sec and x = 0.07 m. (c) Comparison of the velocity of the baseline, RANS-T10A, and RANS-T10B.

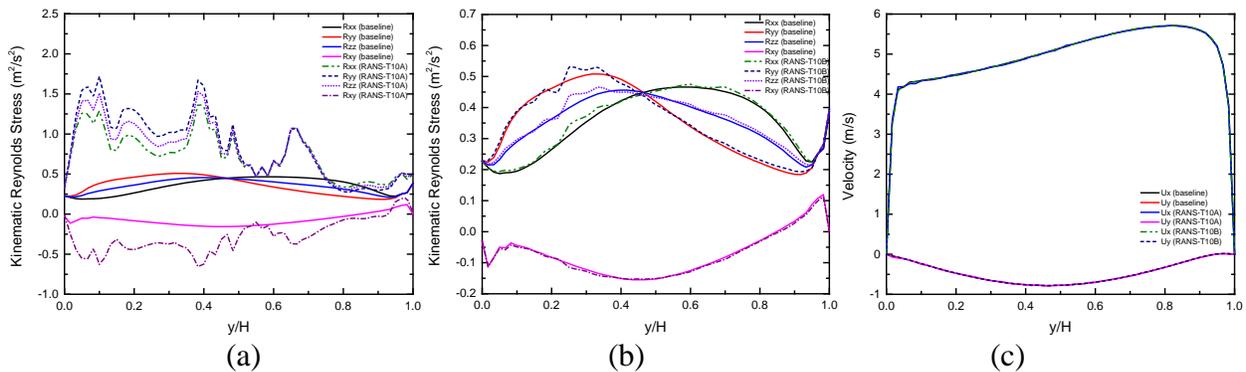

Fig. 15. Comparison of kinematic Reynolds stress (a) between the baseline and RANS-T10A and (b) between the baseline and RANS-T10B at t = 0.01048 sec and x = 0.07 m. (c) Comparison of the velocity of the baseline, RANS-T10A, and RANS-T10B.

### 6.2.1. Visualization of the coverage of flow features in applications by FFCM

We can use flow features coverage mapping (FFCM) to quantify the coverage of flow features in training datasets. Fig. 16 depicts FFCM for RANS-T10A and RANS-T10B at $t = 0.01012$ sec. For RANS-T10A, we compare Fig. 16(a) to Fig. 3(a). Flow features in Fig. 16(a) show different distributions than the features in Fig. 3(a). The discrepancy between two figures can be quantified by the Euclidean distance which is 35.62. The result indicates that T10A dataset is insufficient to cover the transient details in Fig. 12(a). For RANS-T10B, we compare Fig. 16(b) to Fig. 4(a). The two figures have similar distributions since the distance is 4.05 that is much smaller than the distance by RANS-T10A. The result indicates that T10B sufficiently covers the transient details so that RANS-T10B agrees with the baseline in Fig. 12(b).

Fig. 17 shows FFCM for RANS-T10A and RANS-T10B at $t = 0.01048$ sec which is outside of the training domain. Fig. 17(a) shows FFCM for RANS-T10A, and the result is dissimilar to Fig. 3(a) which is FFCM by training data, T10A. Fig. 17(b) depicts that the mapping for RANS-T10B deviates from Fig. 4(b) because the simulation is outside of the training domain. However, the distance is 10.57 which is still smaller than RANS-T10A results. Fig. 15 shows that the performance of RANS-T10B is better than the performance of RANS-T10A.

Table 3 summarizes the distances between distinct FFCM. Table 3 indicates that the distance between RANS-T10B and T10B is much smaller than the distance between RANS-T10A and T10A. The result implies that T10B covers more transient details than T10A. Therefore, RANS-T10B shows good predictive capabilities in Fig. 12-Fig. 15. The analysis by FFCM indicates that RANS-DL can make inferences from training data for prediction when training data sufficiently cover the physics in applications.

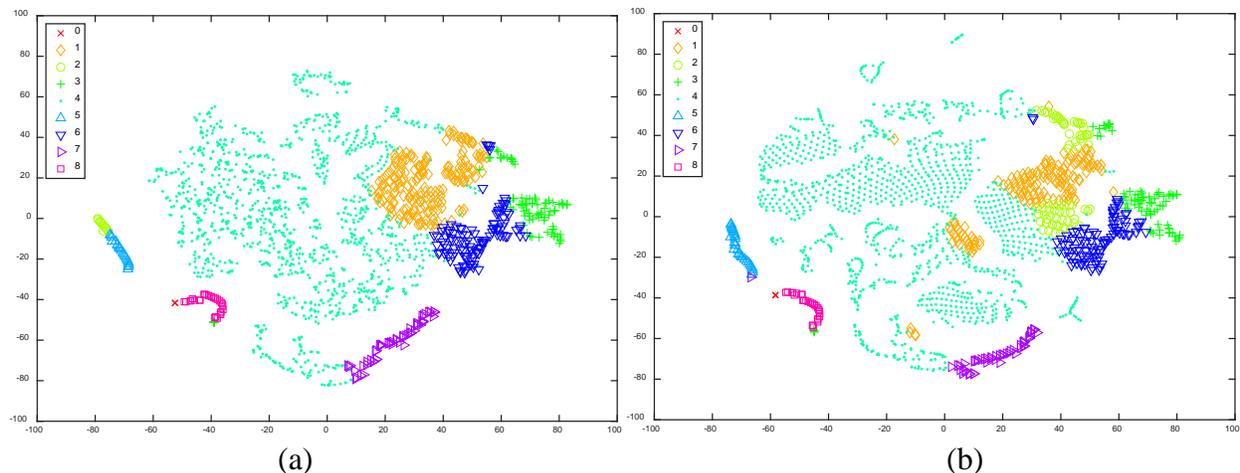

(a)     (b)

Fig. 16. Visualization of flow features coverage mapping (FFCM) using t-SNE for (a) RANS-T10A and (b) RANS-T10B at t = 0.01012 sec. The flow features are clustered by k-means clustering with variously labeled colors.

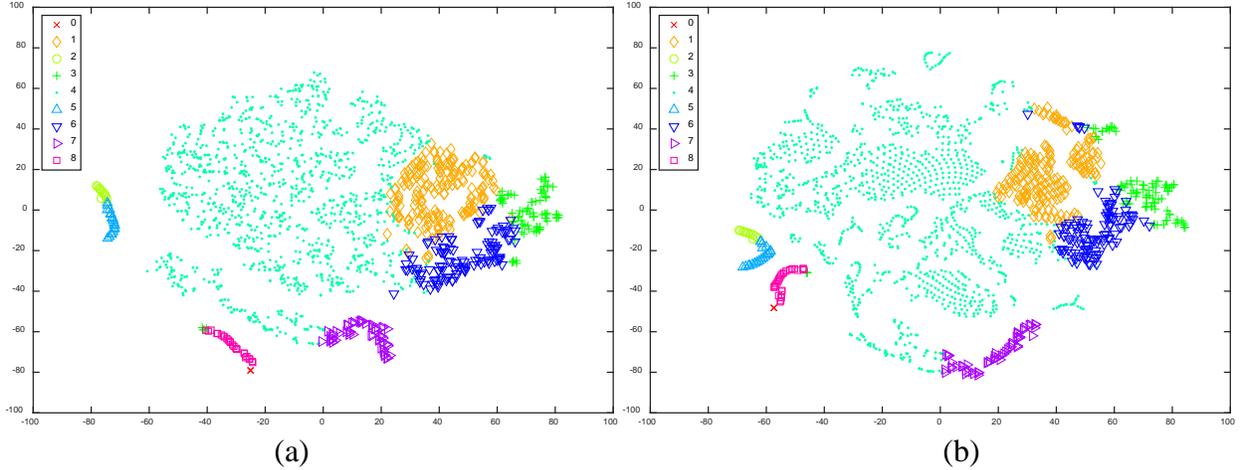

Fig. 17. Visualization of flow features coverage mapping (FFCM) using t-SNE for (a) RANS-T10A and (b) RANS-T10B at t = 0.01048 sec. The flow features are clustered by k-means clustering with variously labeled colors.

Table 3. Summary of Euclidean distances between different FFCM.

| Simulation time (sec) | $d$ between RANS-T10A and T10A (0.01 sec) | $d$ between RANS-T10B and T10B |
|---|---|---|
| 0.01012 | 35.62 | 4.05 |
| 0.01048 | 38.91 | 10.57 |

### 6.2.2. Evaluation of RANS-DL using half of the solver time step

This task is formulated to solve RANS-T10B using half of the solver time step ($1.2 \times 10^{-5}$ sec). Fig. 18 illustrates the comparison between RANS-T10B and the baseline for kinematic Reynolds stress and velocities at three times: 0.010096, 0.01024, and 0.010384 sec. When the solver time step is reduced, DL-based Reynolds stress is not sufficiently trained by those transient conditions. RANS-T10B cannot reproduce the identical solutions as the baseline. However, when RANS-T10B predicts flow transients in the training domain, the discrepancy to the baseline is still smaller than the errors in extrapolation domains. The results indicate that DL-based Reynolds stress can make inferences from the training data.

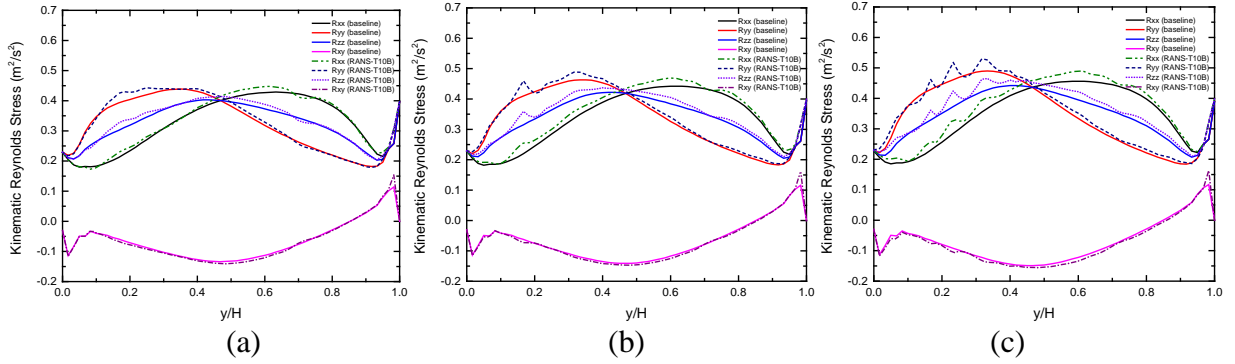

Fig. 18. Comparison of kinematic Reynolds stress at x = 0.07 m between the RANS-T10B and baseline at t = (a) 0.010096, (b) 0.01024, and (c) 0.010384 sec with the solver time step size equal to 1.2x10-5 sec.

### 6.2.3. Evaluation of RANS-DL by perturbing the inlet velocity

In this task, we solve RANS-T10B with ± 10% perturbations of the inlet velocity. Fig. 20 and Fig. 21 show the comparison between RANS-T10B and the baseline with inlet velocities, 11 and 9 m/s. Distinctions between RANS-DL and the baseline are expected since DL-based Reynolds stress is not trained under these two conditions.

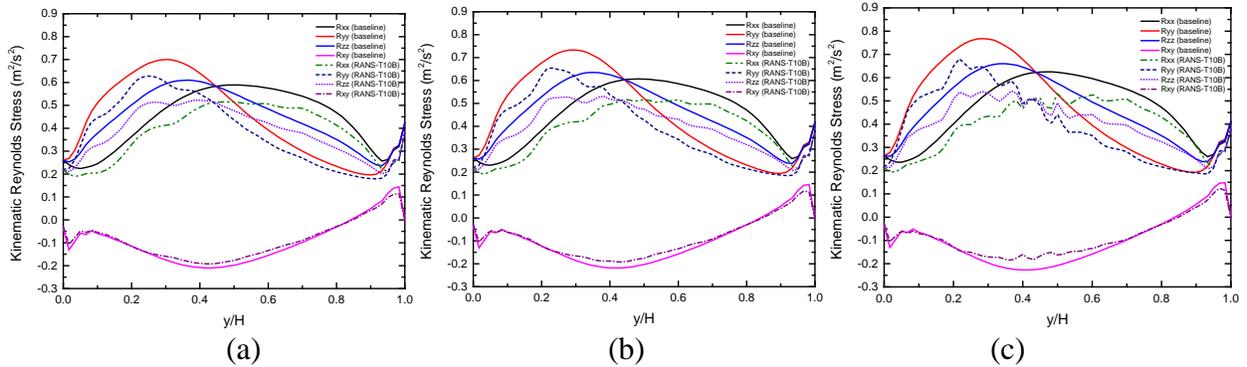

(a) (b) (c)

Fig. 19. Comparison of kinematic Reynolds stress at x = 0.07 m between the baseline and RANS-T10B at t = (a) 0.01012, (b) 0.01024, and (c) 0.01036 sec with the inlet velocity equal to 11 m/s.

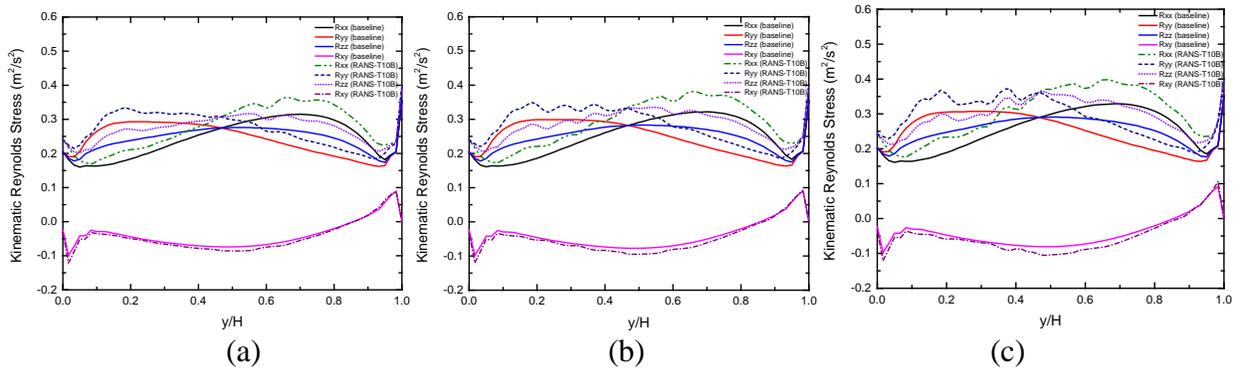

(a) (b) (c)

Fig. 20. Comparison of kinematic Reynolds stress at x = 0.07 m between the baseline and RANS-T10B at t = (a) 0.01012, (b) 0.01024, and (c) 0.01036 sec with the inlet velocity equal to 9 m/s.

### 6.3. Evaluation of the performance of using Type II ML with transient data

The last task is formulated to investigate whether Type II ML can bring RANS-DL to the quasi-steady state from a transient state. Fig. 22 sketches the streamwise velocity field at the quasi-steady state as the reference solution. Fig. 23(a) illustrates the simulations with the various start time ranging from $t = 0.06$ sec to $t = 0.4$ sec. Fig. 23(b) gives the results at $t = 1$ sec. The results reveal that Type II ML can take RANS simulations to the quasi-steady state if initial states are close to the reference solution. Otherwise, RANS-DL by Type II ML can lead to physically unstable solutions.

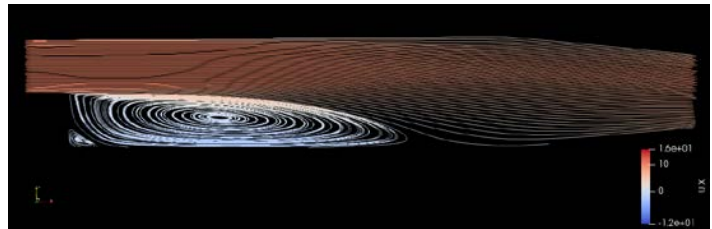

Fig. 21. Streamwise velocity field for the quasi-steady state by the baseline solution at t = 1 sec.

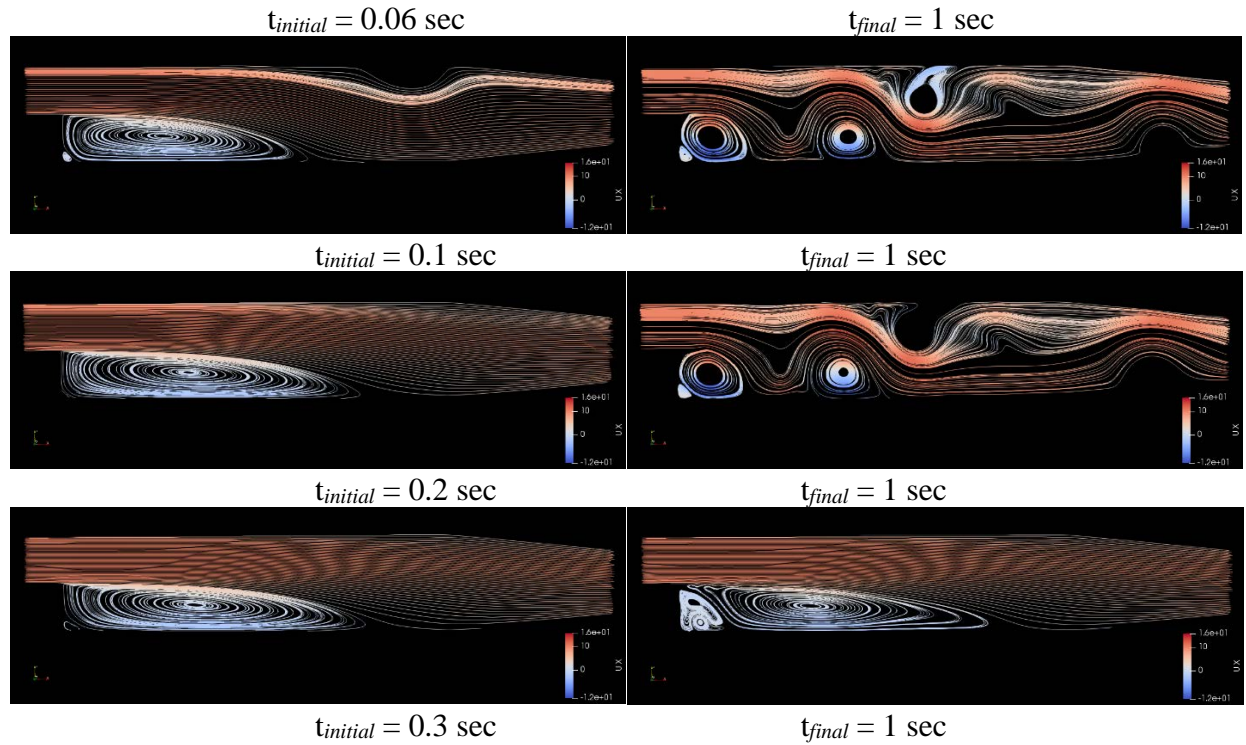

$t_{initial} = 0.06$ sec            $t_{final} = 1$ sec

$t_{initial} = 0.1$ sec             $t_{final} = 1$ sec

$t_{initial} = 0.2$ sec             $t_{final} = 1$ sec

$t_{initial} = 0.3$ sec             $t_{final} = 1$ sec

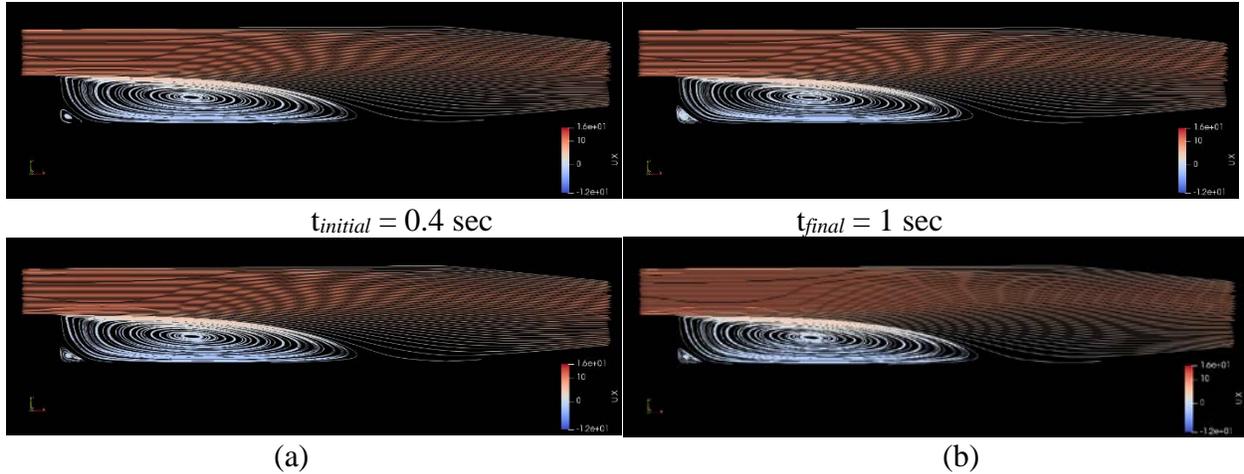

(a)                          (b)

Fig. 22. (a) Initial streamwise velocity field at different transient steps by the baseline solutions. (b) Final streamwise velocity field at t = 1 sec by the RANS model with the fixed field of the Reynolds stress from the quasi-steady-state solution.

Fig. 24(a) shows MSE analysis for RANS simulation with the initial state at $t = 0.06$ sec. The MSE is calculated by evaluating the difference ($dt^n$) between the solutions from two consecutive time steps ($t^{n-1}$ and $t^n$). When the initial state is far from the quasi-steady state, Type II ML leads to physically unstable solutions. Fig. 24(b) depicts the result by using the initial condition at $t = 0.6$ sec. Since the velocity field is close to the reference value, the solution can reach the quasi-steady state. Fig. 25 presents initial MSEs and final MSEs by comparing the reference solution to RANS simulations with distinct start times given in Table 4. Fig. 25 indicates that case 6 is the threshold that allows Type II ML to carry solutions from a transient state to the quasi-steady state. It is noted that the initial condition of case 6 is close to the quasi-steady-state solution.

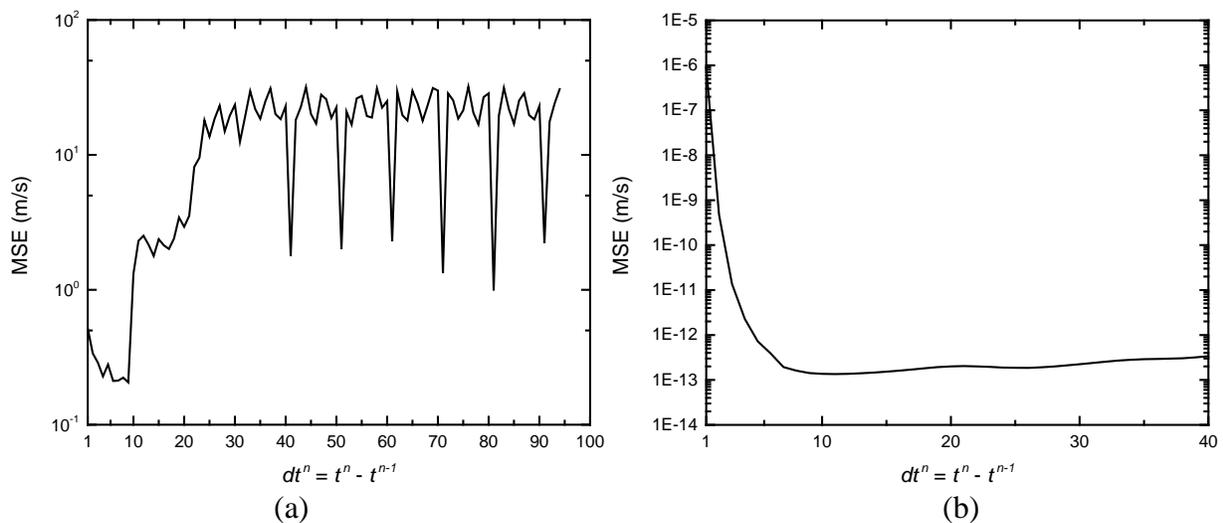

(a)                          (b)

Fig. 23. MSE analysis for showing the solution is (a) unstable when the reference Reynolds stress is injected at t = 0.06 sec and (b) the solution is stable when the reference Reynolds stress is injected at t = 0.6 sec.

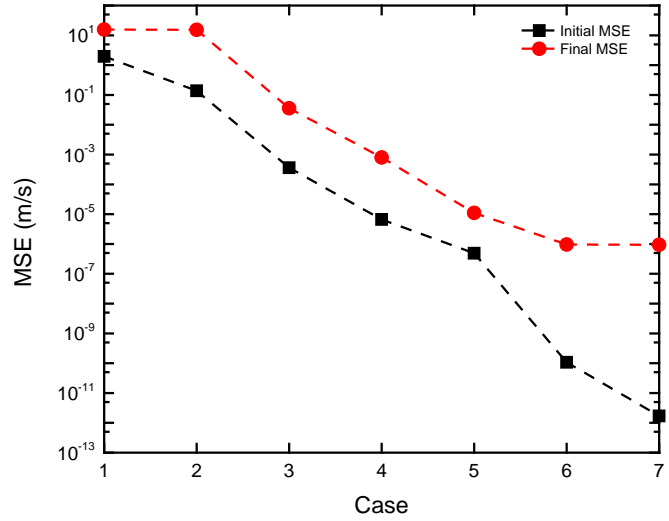

Fig. 24. MSE analysis for searching the threshold discrepancy between the initial transient and reference velocity fields that can bring the transient solution to the quasi-steady state by Type II ML.

Table 4. RANS simulations with different initial states.

| Case | 1 | 2 | 3 | 4 | 5 | 6 | 7 |
|---|---|---|---|---|---|---|---|
| Start time (sec) | 0.06 | 0.1 | 0.2 | 0.3 | 0.4 | 0.5 | 0.6 |

## 7. Lessons learned

Based on the case study in this work, we observed several properties of Type I and Type II ML. Type I ML can deliver DL-based Reynolds stress to close RANS equations for unsteady flow simulation. Training data only requires spatial derivatives of velocity fields without using time derivative quantities because the time history is embedded in transient data. However, data are required to have sufficient spatiotemporal resolutions to include sufficient transient details that allow DL to discover underlying correlations behind data.

The uncertainty of RANS-DL is accumulated along with simulation time if flow features are in extrapolation domains. This is because the physics is not covered by training data, and the coverage of physics can be quantified by computing the Euclidean distance between two flow features coverage mapping (FFCM). When FFCM shows similar distributions between training and applications, RANS-DL can achieve satisfactory performance in prediction. Therefore, when data is insufficient, DL-based Reynolds stress should not be used to predict flow transients which are far from the training domain.

Type II ML can cause physically unstable solutions when initial states of RANS simulation are far from the quasi-steady-state solution. RANS simulation by Type II ML converges to reference solutions only when initial conditions are close enough to reference solutions. This essence limits the use of Type II ML for unsteady flow simulation.

## 8. Conclusions

The present paper demonstrates data-driven turbulence modeling for transient applications that use RANS equations with DL-base Reynolds stress to reproduce RANS (k-ε) solutions. The case study indicates flow features by first-order spatial derivatives of velocity fields are necessary and sufficient to reconstruct the RANS results.

The goal of using DL-based Reynolds stress is to ensure that RANS-DL is globally extrapolating while local variables are within interpolation domains. The results of analysis suggest that DL-based Reynolds stress requires a substantial amount of training data to ensure the predictive capability. The value of data can be evaluated by flow features coverage mapping, which can quantify the coverage of physics by comparing flow features in training and applications.

## Acknowledgments

The support by the US Department of Energy via the Consortium for Advanced Simulation of Light Water Reactors (CASL), and NEUP Integrated Research Project is gratefully acknowledged. The authors thank NVIDIA Corporation for the Titan Xp GPU used for this research.